\documentclass{JAC2003}
\usepackage{amsmath}
\usepackage{graphicx}
\usepackage{subfigure}
\usepackage{booktabs}
\usepackage{url}
\usepackage{algorithm,algorithmic}
\usepackage{tikz}
\usepackage{suffix}
\newcommand{\opal}{\textsc{OPAL}}

\renewcommand{\epsilon}{\varepsilon} 
\renewcommand{\vec}[1]{{\bf #1}}

\def\vec#1{\mathbf{#1}}

\renewcommand {\Re}{{\rm I \kern-2pt R}}

\begin{document}
\title{THE DARK CURRENT AND MULTIPACTING CAPABILITIES IN \opal:\\ MODEL, BENCHMARKS AND APPLICATIONS}

\author{C. Wang\thanks{ cwang@ciae.ac.cn}, Z. G. Yin, T. J. Zhang, CIAE, Beijing, China\\
A. Adelmann\thanks{ andreas.adelmann@psi.ch}  PSI, CH-5232, Switzerland}

\maketitle

\begin{abstract}
Dark current and multiple electron impacts (multipacting), as for example observed in radio frequency (RF) structures of accelerators,
are usually harmful to the equipment and the beam quality. These effects need to be suppressed to guarantee efficient and stable operation.
Large scale simulations can be used to understand causes and develop strategies to suppress these phenomenas.

We extend \opal, a parallel framework for charged particle optics in accelerator structures and beam lines, with the necessary physics models to efficiently and precisely simulate multipacting phenomenas.\ We added a Fowler-Nordheim field emission model, two secondary electron emission models, developed by Furman-Pivi and Vaughan respectively, as well as efficient 3D boundary geometry handling capabilities. The models and their implementation are carefully benchmark against a non-stationary multipacting theory for the classic parallel plate geometry. A dedicated, parallel plate experiment is sketched.
\end{abstract}

\section{INTRODUCTION \label{intro}}
Dark current and multipacting phenomena have been observed in various accelerator RF structures, e.g.~in electron guns, due to field emission caused by strong accelerating fields \cite{J-H-Han}, and multipacting is also appearing in high-Q RF cavities of cyclotrons \cite{CY, riken}. These phenomena are usually harmful to the equipment and beam quality, as they will cause galvanic etching on the surface of the cavity and thus cause RF breakdown.

Multipacting in cyclotron cavities is a very disturbing phenomenon. The seed electrons will impact the cavity surface, and produce an avalanche of new electrons. Under certain conditions (material, geometry of the RF structure, frequency and level of the electromagnetic field), the secondary emission yield (SEY) coefficient will be larger than one and lead to exponential multiplication of electrons. This kind of discharge will limit the power level, until the surfaces are cleaned through a conditioning process. However, this process is very time-consuming \cite{CY, riken}.

 Large scale dark current and multipacting simulations based on reliable data of surface material, full size geometry model of RF structures and parallel computing capabilities, allow more thorough analysis and a deeper understanding of these phenomena, even in early design stages of RF structures.

To make \opal\ \cite{opal:1} a feasible tool for performing large scale dark current and multipacting simulations, we implement a 3D particle-boundary collision test to minimise particle searching during the tracking process. In a subsequent step we add surface physics models including an analytic Fowler-Nordheim field emission model and two secondary emission models, developed by Furman-Pivi and Vaughan respectively. The above mentioned models and their implementation in \opal\ have been benchmarked against a non-stationary theory \cite{Non}.\ A nano-second time resolved multipacting experiment is ongoing.
\section{MODELS }
\subsection{Geometry Handling}
Testing particle-boundary collisions is crucial to both dark current and multipacting simulations. Since complex 3D geometries are hard to be accurately  parameterized by simple functions, we use triangulated surfaces, which are extracted from a volume mesh generated by GMSH \cite{gmsh}, to represent the complex geometries. Subsequently we can make use of efficient 3D line segment-triangle intersection (LSTI) tests to identify particle-boundary collisions.
\begin{figure}[H]
\centering{
    \resizebox{0.25\textwidth}{!}{\begin{tikzpicture}
\draw[step=1cm,gray,dashed] (-3,0) grid (3,6);
\draw[step=1cm,gray,dashed,fill=gray!30] (-3,3) rectangle (-2,4); 
\draw[step=1cm,gray,dashed,fill=gray!30] (-2,1) rectangle (-1,2); 
\draw[step=1cm,gray,dashed,fill=gray!30] (-2,2) rectangle (-1,3); 
\draw[step=1cm,gray,dashed,fill=gray!30] (-2,3) rectangle (-1,4);
\draw[step=1cm,gray,dashed,fill=gray!30] (-1,1) rectangle (0,2);
\draw[step=1cm,gray,dashed,fill=gray!30] (-1,0) rectangle (0,1);
\draw[step=1cm,gray,dashed,fill=gray!30] (0,0) rectangle (1,1);
\draw[step=1cm,gray,dashed,fill=gray!30] (0,1) rectangle (1,2);
\draw[step=1cm,gray,dashed,fill=gray!30] (1,1) rectangle (2,2);
\draw[step=1cm,gray,dashed,fill=gray!30] (1,2) rectangle (2,3); 
\draw[step=1cm,gray,dashed,fill=gray!30] (1,3) rectangle (2,4);
\draw[step=1cm,gray,dashed,fill=gray!30] (2,3) rectangle (3,4);
\draw[step=1cm,gray,dashed,fill=gray!30] (-3,4) rectangle (-2,5);
\draw[step=1cm,gray,dashed,fill=gray!30] (-3,5) rectangle (-2,6);
\draw[step=1cm,gray,dashed,fill=gray!30] (2,4) rectangle (3,5);
\draw[step=1cm,gray,dashed,fill=gray!30] (2,5) rectangle (3,6);   
\draw[black, thick] (0.,0.5) parabola  ( 2.5,5.5); 
\draw[black, thick] (0.,0.5) parabola  ( -2.5,5.5);
\draw [->,gray] (-1.55,2.5) -- (-1,2.8);
\node[above] (n) at (-1,2.8) {$\vec{n}$};
\draw [<-,gray] (1.0,4.1) -- (2,3.7);
\node[above] (n) at (1,4.1) {$\vec{n}$};
\draw (-1.2,2.1) node[circle,fill=yellow]{};
\draw [->,thick] (-1.2,2.1) -- (-0.3,2.1);
\draw (0,4.1) node[circle,fill=green]{};
\draw [->,thick] (0,4.1) -- (0.3,4.4);
\draw (1.2,3.5) node[circle,fill=red]{};
\draw [<-,thick] (1.8,3.5) -- (1.2,3.5);
\end{tikzpicture}}}
    \caption{Schematic view of the particle-boundary early rejection strategy. The dark
    black line represents the boundary surface, particles are coloured dots with an
    attached momenta arrow. Gray arrows visualize inward normals of the
    boundary.\label{fig:P-B}}
\end{figure}
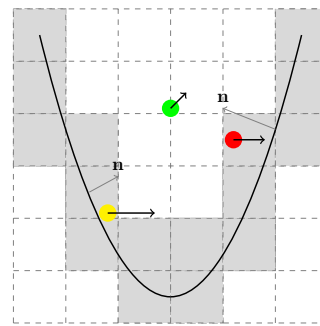
Our LSTI test algorithm is based on \cite{LT}, and the detailed introduction to our implementation can be found in our previous paper \cite{WangHB2010:1}.\
Early rejection strategies skip particles far away from the
boundary and with equal direction of momenta and boundary normal
vectors (see Figure Figure~\ref{fig:P-B})

\subsection{Surface Physics Models}
Electron field emission is a major source of both dark current particles and primary
incident particles in secondary emission. We employ the Fowler-Nordheim (F-N) formula (\ref{eq:units})
to predict the emitted current density
\cite{FN, Feng}
\begin{equation}\label{eq:units}
    J_{FN}(\mathbf{r},t) = \frac{A(\beta E)^2}{\varphi t(y)^2}
                      \exp{\left(\frac{-B v(y)\varphi^{3/2}}{\beta E}\right)}
\end{equation}
where $J_{FN}(\mathbf{r},t)$ stands for emitted electric current density in position
$\mathbf{r}$ and time $t$. The Greek letters $\varphi$ and $\beta$ denote the
work function of the surface material and the local field enhancement factor
respectively. The parameter $E$ is the electric field in the normal direction
of surface. The parameters $A$ and $B$ are empirical constants. The functions
$v(y)$ and $t(y)$ representing the image charge effects \cite{Feng} as a function
of the Fowler-Nordheim parameter $y$ with the following definition \cite{J-H-Han}
\begin{equation}\label{eq:imagecharge}
    y = \sqrt{\frac{e^3}{4\pi\varepsilon}}\frac{\sqrt{\beta E}}{\varphi}
      = 3.795\times10^{-5}\frac{\sqrt{\beta E}}{\varphi} \text{.}
\end{equation}
In our model, we have chosen simpler approximations $v_a(y)$ and $t_a(y)$ for $v(y)$ and $t(y)$ \cite{J-H-Han}:
\begin{eqnarray*}
v_a(y) &=& a-by^2 \\
t_a(y) &\approx& 1 \text{.}
\end{eqnarray*}
where $a$ and $b$ are free parameters to fit the value of $v(y)$. These approximations are valid for a large range of $y$, corresponding to
typical applied electric field ranges in RF guns.

Whenever the normal components of an electric field are strong enough the field
emission current density will be limited by space charge effects \cite{Feng}.
To cover this situation we incorporated the 1D Child-Langmuir law
\begin{align}\label{eq:SpaceCharge}
    J_{SC}(\mathbf{r},t) & =\frac{4\varepsilon_0}{9}\sqrt{2\frac{e}{m}}\left(\frac{V^{3/2}}{d^2}\right)\notag\\
    &
    =\frac{4\varepsilon_0}{9}\sqrt{2\frac{e}{m}}\left(\frac{E^{3/2}}{d^{1/2}}\right)
\end{align}
into our field emission model. $J_{SC}(\mathbf{r},t)$ denotes space charge limited emission
current density in position $\mathbf{r}$ and time $t$, $\varepsilon_0$ the
permittivity in vacuum, $E$ the normal component of electric field on the surface
and $d$ the distance from the position where $E$ is evaluated. Currently we
choose $d$ to be equal to the distance traveled by emitted particles in one
time step, i.e., $d=\frac{\displaystyle eE\Delta{t}^2}{\displaystyle 2m_0}$ where $\Delta{t}$ is simulation
time step. In each time step, the emitted current density $J(\mathbf{r},t)$ for each surface triangle will be the smaller one of $J_{FN}(\mathbf{r},t)$ and $J_{sc}(\mathbf{r},t)$,
\begin{equation}
J(\mathbf{r},t)=min\{J_{FN}(\mathbf{r},t), J_{sc}(\mathbf{r},t)\}.
\end{equation}

We implemented two secondary emission models. The first one is a
phenomenological model developed by M. A. Furman and M. Pivi \cite{Furman-Pivi}, generating various generations of secondary electrons: {\em true secondary}, {\em rediffused} or {\em backscattered}.

The other secondary emission model is based on a secondary emission yield formula developed by Vaughan \cite{Vaughan, VaughanRv, FS}:
\begin{subequations}
\label{Vaughanall}
\begin{eqnarray}
\delta(E,\theta)&=&\delta_0,\ \text{for}\ v \le 0 \label{eq:VaughanA}
\\
    \delta(E,\theta)&=&\delta_{max}(\theta)\cdot (v e^{1-v})^k,\ \text{for}\ v \le 3.6 \label{eq:VaughanB}
\\
\delta(E,\theta)&=&\delta_{max}(\theta)\cdot 1.125/v^{0.35},\ \text{for}\ v > 3.6 \label{eq:VaughanC}
\end{eqnarray}
\end{subequations}
where
\begin{eqnarray*}
v=\frac{\displaystyle E-E_0}{\displaystyle E_{max}(\theta)-E_0},
\end{eqnarray*}
\begin{eqnarray*}
k=0.56,\ \ \text{for}\ v<1,
\end{eqnarray*}
\begin{eqnarray*}
k=0.25,\ \ \text{for}\ 1<v\le{3.6},
\end{eqnarray*}
\begin{eqnarray*}
\delta_{max}(\theta)=\delta_{max}(0)\cdot (1+k_{\theta}\theta^2/2\pi),
\end{eqnarray*}
\begin{eqnarray*}
E_{max}(\theta)=E_{max}(0)\cdot (1+k_E\theta^2/2\pi).
\end{eqnarray*}
The secondary emission yield value for an impacting electron with energy $E$ and incident angle $\theta$ w.r.t the surface normal is denoted as $\delta(E,\theta)$. Parameter $k_{\theta}$ and $k_E$ denote the dependence on surface roughness. Both
should be assigned a default value of 1.0, which appears appropriate for typical dull surfaces in a working tube environment.
Lower values down to zero or higher values, up to about 2.0, are only valid for specific cases \cite{Vaughan}. $E_{max}(0)$ is the impacting energy when the incident angle is zero and secondary yield reaches its maximum value. $E_0$ is an adjustable parameter to make the first crossover energy at which the secondary yield equals to 1 be fitted to the experiment data \cite{FS}.

\subsection{Implementation within \opal}
The above models are implemented in the object-oriented parallel ESPIC code \opal \cite{opal:1}.\


Statistical data, such as particle position, momentum and particle type (primaries, field emitted electrons or true secondaries), are stored in the H5hut \cite{H5hut:1} file format. In a post processing step, the data can be converted into legacy VTK formatted files \cite{vtk:1} and processed by standard visualisation tools.

Users can customise the dark current and multipacting models in the MAD like
input file of \opal. The input file format and examples can be found in \opal\ user guide \cite{opal:1}.

Motivated by the fact that the number of simulation particles may grow exponentially, spanning a range of ten orders of magnitude, a particle re-normalisation technique is implemented.
In each electron impact event, instead of emitting the real number of simulation particles predicted by secondary emission models, this re-normalization approach emits only one particle and a scaled charge $Q_s$. Where $Q_s = Q_{incident} \times \delta$, the incident particle charge multiplied by the SEY value $\delta$. This approach is an accurate representation of the secondary emission model which can be observed in the following parallel plate benchmarking cases.

\section{BENCHMARKS}
\subsection{Benchmark Against the None-stationary Theory}
\begin{figure}[H]
\centering{
    \resizebox{0.4\textwidth}{!}{\scalebox{0.7}{
\begin{tikzpicture}
\usetikzlibrary{arrows}
\draw [<->,thick] (0,0.8) node (zaxis) [above] {$\mathbf{z}$}
        |- (0.8,0) node (yaxis) [right] {$\mathbf{y}$};
\draw [->,thick] (0,0) -- (-0.5656,-0.5656) node (xaxis) [above] {$\mathbf{x}$};

\draw (-3.5,-1) -- (2.5,-1);
\draw (2.5,-1) -- (5,3);
\draw (0.5,3) -- (5,3);
\draw (-3.5,-1) -- (0.5,3);
\draw [<-] (-3.5,-1.05) -- (-3.5,-2.1) node (d) [left,below] {$\mathbf{d}$};
\draw [<-] (-3.5,-3.75) -- (-3.5,-2.7);
\draw (-3.5,-3.8) -- (2.5,-3.8);
\draw (2.5,-3.8) -- (4.5,-0.5);
\draw (-0.0,-0.5) -- (4.5,-0.5);
\draw (-3.5,-3.8) -- (0.0,-0.5);

\path[draw=black] (3.3,0.5) circle (2pt);
\path[draw=black,thick] (5.5,-1) circle (0.4);
\draw [] (3.3,0.5) arc (90:37:2.9);
\path[draw=black] (3.3,-2.5) circle (2pt);
\draw [] (3.3,-2.5) arc (-90:-34:2.6);
\draw [thick] (5.1,-1) sin (5.3,-0.9) cos (5.5,-1) sin (5.7,-1.1) cos (5.9,-1) sin (5.9,-1);
\node[above=7pt,right=12pt] (I) at (5.5,-1) {$\vec{E}=-\mathbf{\hat{z}}\frac{V_0}{d} \sin \omega t$};
\end{tikzpicture}
}}}
\caption{The geometry of the analytical model.\label{fig:sk}}
\end{figure}
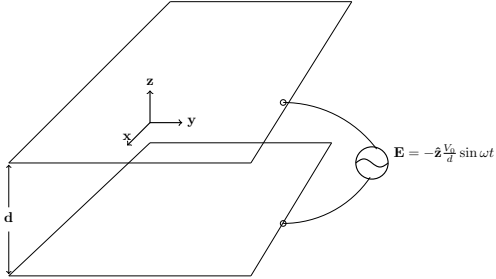

The theory we use to benchmark the described models is restricted to the case of the simple plane-parallel model of multipactor. We consider a spatially homogeneous and time harmonic RF field in between and directed perpendicular to the plates i.e.,
\begin{equation}
\vec{E}(\vec{z},t) = -\mathbf{\hat{z}}E_0\sin \omega t=-\mathbf{\hat{z}}\frac{V_0}{d}\sin \omega t
\end{equation}
as shown in Fig.\ \ref{fig:sk}. Electrons are assumed to oscillate between two parallel plates separated by a distance $d$. The $x$ and $y$ dimensions of the plates are assumed to be infinite. The impact of an electron with the plates is accompanied by a secondary emission yield, which depends on the energy and angle of the primary electron and material property of the multipactor.

Assuming electrons are initially generated on the surface of the lower parallel plate, i.e., $z=0$ at $t=t_0$ according to Fig.\ \ref{fig:sk}, then the equation of motion reads
\begin{equation}
\frac{d^2z}{dt^2} = -\frac{e}{m} E_0\sin\omega t= - \frac{e}{m}\frac{V_0}{d}\sin\omega t \label{scalar}
\end{equation}
with $V_0$ being  the peak voltage amplitude between the parallel plates.

Integrating the equation \eqref{scalar} w.r.t variable $t$, following \cite{PP}, i.e., substituting the initial condition $\displaystyle \frac{dz}{dt}|_{t=t_0}=v_{0}$, $z|_{t=t_0}=0$, and using normalised variables:  $v_{\omega}=eV_0/m\omega d $, $\lambda=\omega d/v_{\omega}$, $u=v_{0}/v_{\omega}$, $\omega t_0=\varphi_0$, we obtain the scaled velocity and absolute position of electrons before they impact as:
\begin{equation}
\begin{split}
z=&-\frac{d}{\lambda}\sin\omega t+\frac{d}{\lambda}(u+\cos\varphi_0)\omega t\\
&+\frac{d}{\lambda}\sin\varphi_0-\frac{d}{\lambda}(u+\cos\varphi_0)\varphi_0.\label{po}
\end{split}
\end{equation}
If we define $\omega t=\varphi$, $\xi=\omega z/v_{\omega}$ and $\tau=\varphi-\varphi_0$, equation \eqref{po} can be rewritten as:
\begin{equation}
\xi(\varphi,\varphi_0,u) = (u+\cos \varphi_0)\tau+\sin \varphi_0 - \sin (\varphi_0+\tau).\label{nposition}
\end{equation}

\begin{figure}[H]
\begin{center}
\scalebox{0.7}{
\begin{tikzpicture}
\usetikzlibrary{arrows}
\draw[fill=gray!60,dashed] (-3,0) rectangle (6,0.3);
\draw[fill=gray!60,dashed] (-3,-4) rectangle (6,-3.7);
\draw[very thick] (-3,-3.7) .. controls (-2.5,-3.5) and (-1.5,-2) .. (-1,0);
\draw[very thick] (-1,0) .. controls (-0.7,-0.4) and (-0.2,-0.4) .. (0.1,0);
\draw[very thick] (0.1,0) .. controls (0.8,-2.4) and (2.2,-1.6) .. (3.2,-3.7);
\draw[very thick] (3.2,-3.7) .. controls (3.4,-3.1) and (3.7,-3.0) .. (4.2,-3.7);
\draw (5.5,-3.7) node (d) [above] {$\mathbf{D}$};
\draw (5.5,0) node (u) [below] {$\mathbf{U}$};
\draw (-2.0,-2.5) node (ds1) [right] {$\mathbf{ds}$};
\draw (-0.45,-0.25) node (ss1) [below] {$\mathbf{ss}$};
\draw (1.5,-2) node (ds2) [below] {$\mathbf{ds}$};
\draw (3.6,-3.2) node (ss2) [above] {$\mathbf{ss}$};
\draw [<-,thick] (-3.,-3.7) -- (-3,-1.85) node (d) [left] {$\mathbf{\lambda}$};
\draw [<-,thick] (-3,0) -- (-3,-1.85);
\end{tikzpicture}
}
\end{center}
\caption{ A full scenario of electron's trajectories between parallel plates, including both double surface ($ds$) and single surface ($ss$) impacts \cite{Non}\label{fig:ss-ds}.}
\end{figure}
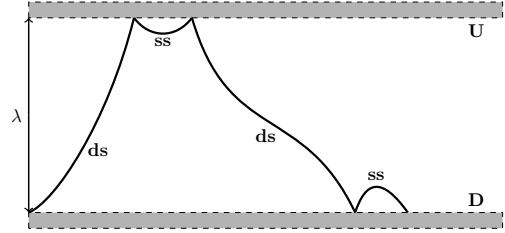

A non-stationary statistical theory, originated from a stationary statistic theory \cite{ST}, for multipactor introduced by Anza et al.\ \cite{Non}, gives a more realistic scenario than previous multipacting theories, since it considers the random nature of the electron emission velocity and models both double and single surface impacts, as sketched in Fig.\ \ref{fig:ss-ds}.

The basic idea of the non-stationary theory can be summarised as follows: the initial velocity $u$ of emitted particles is a random variable, the solution of equation \eqref{nposition} with respect to time $\tau$ is the joint probability that an electron released at phase $\varphi_s$ impacts at the opposite wall, separated by $\lambda$, in a transit time $\tau$. As long as we know the probability density function (PDF) of the initial velocity $u$, which usually is a thermal distribution, then the joint PDF can be derived according to the rule of change of variable in probability theory.

Following the definitions in \cite{Non}, each plate will be denoted as $D$ and $U$ for the boundary condition $\xi=0$ (down) and $\xi=\lambda$ (up), respectively. Double and single surface impacts with $D-U$ or $U-D$ and $D-D$ or $U-U$ trajectories will be denoted as $ds$ and $ss$ respectively. Other definitions for the non-stationary theory are given in \cite{Non}, and listed here in Table \ref{my_table} for convenience.
\begin{table}[h]\footnotesize
\caption{Non-stationary theory definitions.}
\centering
  \label{my_table}
  \begin{tabular}{p{4.7cm} r }
    \hline
\hline
    Impact rate (electrons/radian) in plate $U/D$ at phase $\varphi$ & $I_{U/D}(\varphi)$ \\
    Emission rate (electrons/radian) in plate $U/D$ at phase $\varphi$ & $C_{U/D}(\varphi)$ \\
    Number of electrons at time $\varphi$ & $N(\varphi)$ \\
    Probability density that an electron starting at plate $U/D$,
    with starting phase $\varphi$, experiences a double/single
    Surface impact in a transit phase $\tau$ & $G_{ds/ss,U/D}(\tau|\varphi)$\\
    Secondary emission yield of an electron starting at plate $U/D$, with starting
    phase $\varphi$, experiences a double/single surface impact
    in a transit phase $\tau$ & $\delta_{ds/ss,U/D}(\tau|\varphi)$\\

    \hline
 \hline
  \end{tabular}
 \end{table}

The electron emission and impact rates in each plate can be described by the joint PDF and the secondary emission yield coefficient, both of which are functions of initial velocity $u$, initial phase $\varphi_s$ and time $\tau$ at which particles hits the plates. Details of constructing the joint PDF and integrating the emission rates and impact rates can be found in \cite{Non}. The number of electrons between the parallel plates at phase $\varphi$ then can be integrated as:
\begin{flalign}
N(\varphi)=&\int_0^\varphi \left(C_U(\varphi')+C_D(\varphi')-I_U(\varphi')-I_D(\varphi')\right)\mathrm{d}\varphi'\label{npdef}.
\end{flalign}

We use the secondary emission yield curve of copper provided in \cite{Furman-Pivi} to benchmark the Furman-Pivi model and use the secondary emission yield curve of silver given in \cite{Non} to benchmark Vaughan's  secondary emission model.

\begin{figure}[hbtp]
\begin{center}
\includegraphics[width=0.55\textwidth]{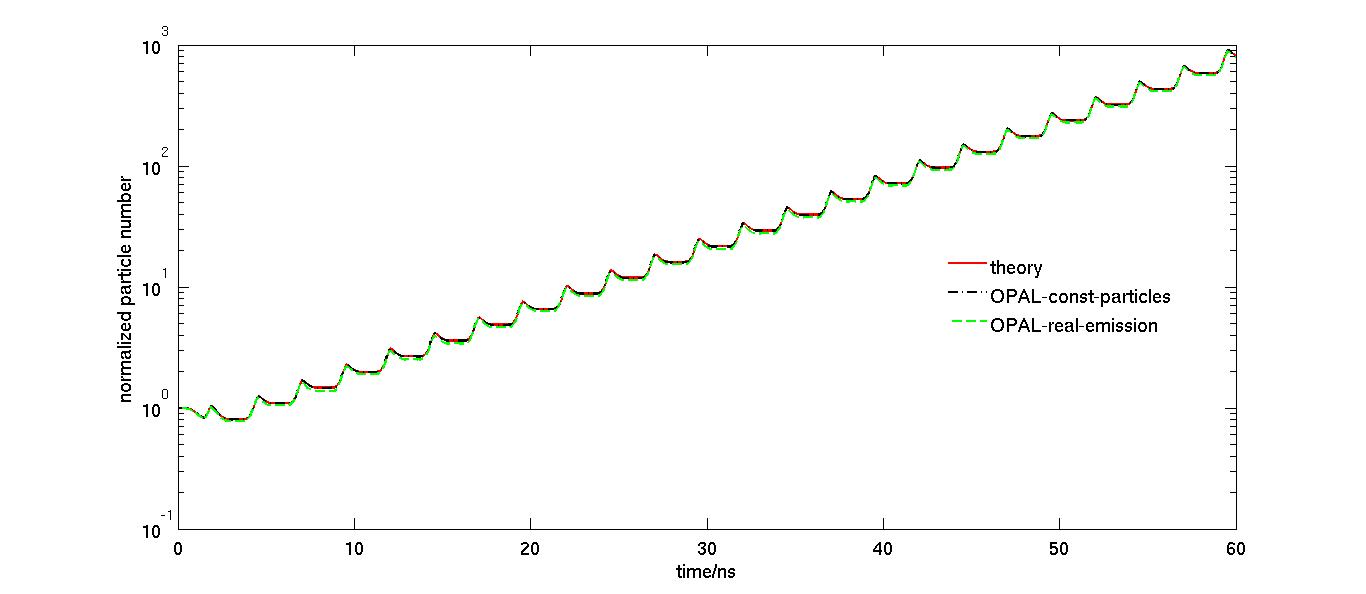}
\end{center}
\caption{Time evolution of the electron density predicted by the theoretical model and the \opal\ simulation. Furman-Pivi's secondary emission model with and without re-normalization is used. Model parameter are: $f=200$ MHz, $V_0=120$ V, $d=5$ mm. \label{fig:results}}
\end{figure}

\begin{figure}[hbtp]
       \begin{center}
       \includegraphics*[width=0.55\textwidth]{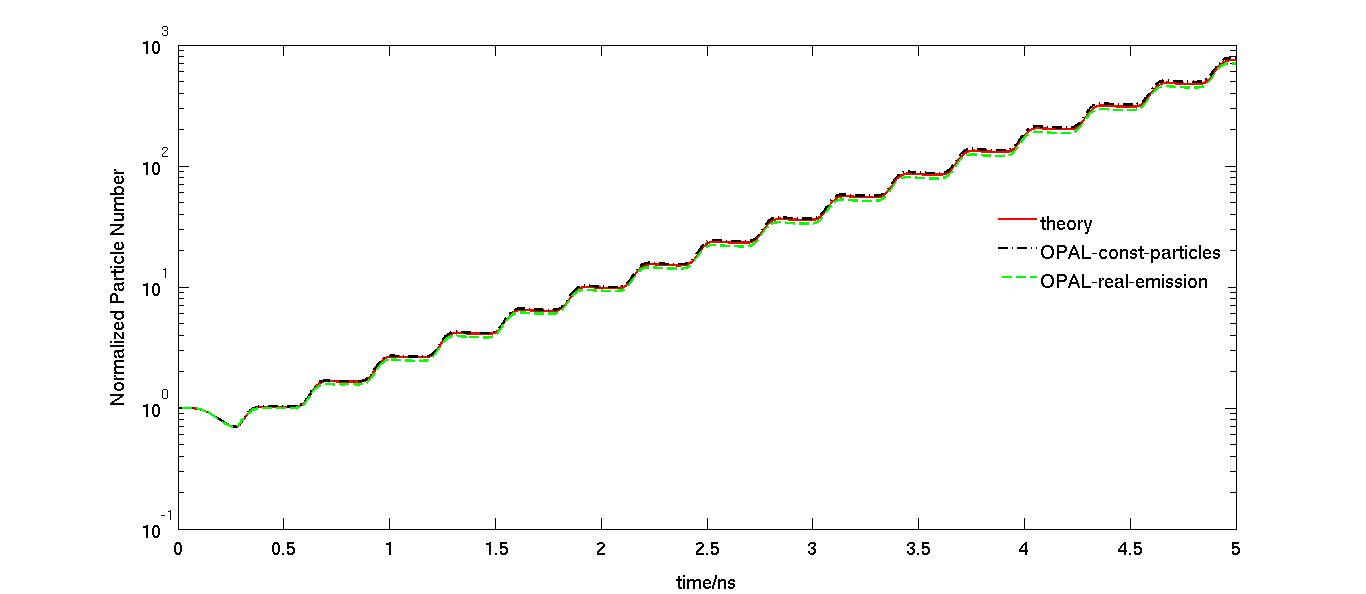}
   \end{center}%
   \caption{Time evolution of the electron density predicted by the theoretical model and the \opal\ simulation.  Vaughan's secondary emission model with and without re-normalization is used. Model parameter are: $f=1640$ MHz, $V_0=120$ V, $d=1$ mm. \label{fig:multi_silver}}
 \end{figure}

The initial particles are equally distributed between both plates. The velocity of emitted particles both in the theory and in the benchmark simulation follows a Maxwellian distribution \cite{Non}.
We have chosen different surface material SEY curves, voltages, gap distances,
frequencies to benchmark against both Furman-Pivi's model and Vaughan's model. 
Fristly we using $f=200$ MHz, $V_0=120$ V, $d=5$ mm and the material of the multipactor is copper. The results with  Furman and Pivi's secondary emission model matches the theoretical model very well, as shown in Fig. \ref{fig:results}.
 The Vaughan's model has been benchmarked with silver SEY data. Again very
 good agreement of the model and simulation can be observed in Fig.\
 ~\ref{fig:multi_silver}.

\section{APPLICATIONS}
Multipacting phenomenas of the CYCIAE-100  H$^-$ AVF cyclotron under construction at the China Institute of Atomic Energy (CIAE)
 \cite{Zhang20084117} are investigated.

We have chosen two different SEY curves of copper, with and without surface treatment, both in agreement with equation \eqref{Vaughanall}.

\begin{figure}[H]
   \centering
  \includegraphics*[width=0.99\linewidth,angle=0]{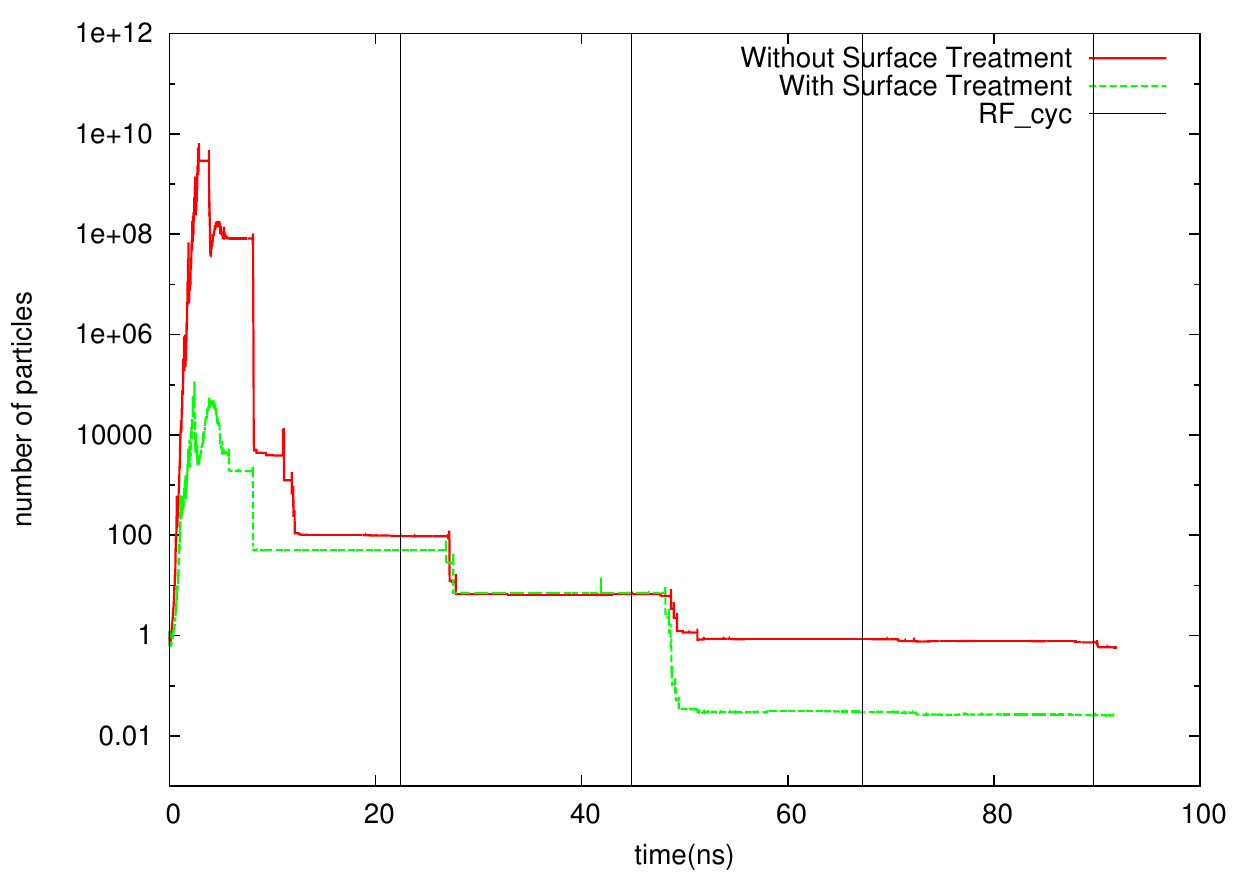}
   \caption{The time evolution of particle numbers in both with and without surface treatment cases, the phase lag of both cases are 0 degree.}
   \label{fig:seydiff}
\end{figure}

Figure \ref{fig:seydiff} shows the time evolution of the electron density. Multipacting has been observed in both cases, with and without surface treatment  within one RF cycle. The electron multiplication without surface treatment is 5 orders of magnitude larger than in the case with surface treatment.

\begin{figure}[H]
\begin{center}
\includegraphics[width=0.5\textwidth]{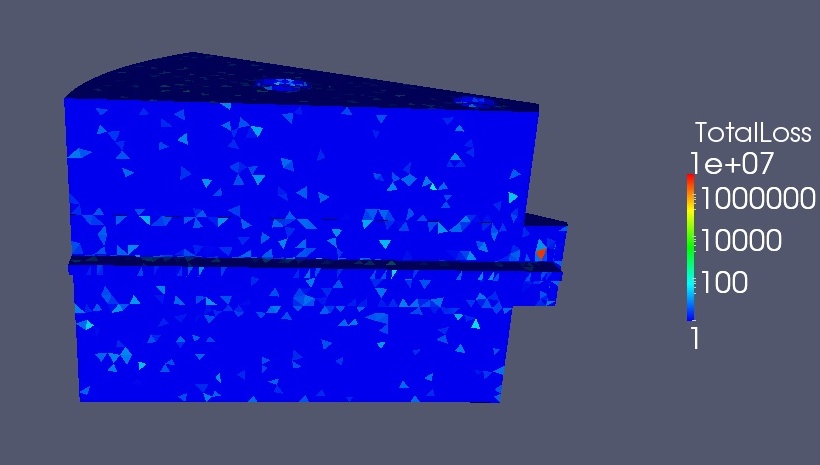}
\end{center}
\caption{Simulated hot spot of RF cavity of CYCIAE-100 cyclotron.\label{fig:hot}}
\end{figure}

For visualising the positions where multipacting happens, \opal\ will dump the impact position and current of incident particles into a specified file using the
efficient parallel H5hut file format \cite{H5hut:1}. The simulated hot spot of the RF cavity of CYCIAE-100 cyclotron is shown in figure \ref{fig:hot}.

 The Dark current module of \opal\ has also been used to study the Dark current of the CTF3 gun in SwissXFEL project \cite{dark_usr}.
\section{CONCLUSION AND OUTLOOK}
RF components with arbitrary, closed structures, like electron guns and cyclotron resonators can be modelled in \opal\ to numerically study dark current and multipacting phenomena. Detailed benchmarks against the non-stationary multipacting theory, enables predictive numerical studies of multipacting phenomena.
Post processing utilities, like the hot spot visualisation and time evolution of particle distribution, are useful to locate the multipacting zones.
Further multipacting simulations are planed for the CYCIAE-100 cyclotron cavity on different RF power levels to better understand the multipacting behaviour of the cavity during the RF conditioning process. The aim is to shorten the time consuming RF conditioning process, by localised surface treatment (painting)  predicted by \opal\ simulations.

\subsection{Time Resolved Multipacting Experiment}
A dedicated nano-second time resolved multipacting experiment is ongoing to benchmark the models.
This dedicated experiment is using a $\lambda /4$ transmission line resonator, which works around 73 MHz. A sinusoidal electrical field will be established between parallel plates, in the conductive end of the resonator. An electron pickup is mounted through a hole in the centre of the ground plate to collect the multipacting electrons. It is expected that the collected multipacting current will have the same pattern as predicted by the none-stationary theory and the  \opal\ model. The experiment is still ongoing, detailed results will be published in a forthcomming  paper.
\section{ACKNOWLEDGMENTS}
The authors thank the Accelerator Modelling and Advanced Simulation (AMAS) and the members of the BRIF department of China Institute of Atomic Energy (CIAE) for many fruitful discussions.  This work was performed on the {\it felsim} cluster at the Paul Scherrer Institut and the PANDA cluster at CIAE.


\begin{thebibliography}{99} 

\bibitem{J-H-Han}
J. H. Han, ``Dynamics of Electron Beam and Dark Current in Photocathode RF Guns,''
PhD thesis, Universit{\"a}t Hamburg, Germany, 2005, \texttt{http://www-library.desy.de/\\*preparch/desy/thesis/desy-thesis-05-045.pdf}
\bibitem{CY} P. K. Sigg, ``Reliability of High Beam Power Cyclotron RF-Systems at PSI,''Workshop on Utilization and Reliability of High Power Proton Accelerators, Mito, Japan, 1998, \texttt{http://rf.web.psi.ch/files/proceedings/1998/\\*JAERI98/PaperNEA98.pdf}
\bibitem{riken} N. Sakamoto et al., ``RF-System for the RIBF Superconducting Ring Cyclotron,''  CYCLOTRONS'2007, Giardini Naxos, Italy, 2007.
\bibitem{opal:1}
A. Adelmann et al., ``The OPAL (Object Oriented Parallel Accelerator Library) Framework,'' Paul Scherrer Institut, PSI-PR-08-02, 2008.
\bibitem{Non} S. Anza et al., Phys. Plasmas. 17 (2010) 6 062110.
\bibitem{gmsh} C. Geuzaine and J. F. Remacle, International Journal for Numerical Methods in Engineering. 79 (2009) 11 1309.
\bibitem{LT} D. Sunday, \texttt{http://softsurfer.com/Archive/\\*algorithm\_0105/algorithm\_0105.htm}.
\bibitem{WangHB2010:1} C. Wang et al., ``A Field Emission and Secondary Emission Model in OPAL,'' HB2010, Morschach, Switzerland, 2010, MOPD55.
\bibitem{FN} R. H. Fowler and L. Nordheim, Royal Society of London Proceedings Series A. 119 (1928) 173-181.
\bibitem{Feng} Y. Feng and J. P. Verboncoeur, Phys. Plasmas. 13 (2006) 7 073105.
\bibitem{Furman-Pivi} M. A. Furman and M. T. F. Pivi, Phys. Rev. ST Accel. Beams. 5 (2002) 12 124404.
\bibitem{Vaughan} J. R. M. Vaughan, IEEE Transactions on Electron Devices. 36 (1989) 9 1963.
\bibitem{VaughanRv} J. R. M. Vaughan, IEEE Transactions on Electron Devices. 40 (1993) 4 830.
\bibitem{FS} C. Vicente et al., ``FEST3D - A Simulation Tool for Multipactor Prediction,'' Proc. of MULCOPIM 2005, Noordwijk, the Netherlands, 2005.
\bibitem{H5hut:1} A. Adelmann et al., \texttt{http://h5part.web.psi.ch/}.
\bibitem{vtk:1} ``File Formats for VTK Version 4.2,'' \texttt{http://www.vtk.org/VTK/img/file-formats.pdf}.
\bibitem{PP} A. Kryazhev et al., Phys. Plasma. 9 (2002) 11 4736.
\bibitem{ST} N. K. Vdovicheva et al., Radiophysics and Quantum Electronics. 47 (2004) 8 580.
\bibitem{Zhang20084117} T. J. Zhang et al., NIM B. 266 (2008) 19 4117-4122.
\bibitem{seycurve} I. Bojko et al., ``The Influence of Air Exposures and Thermal Treatments on the Secondary Electron Yield of Copper,'' LHC Project Report, CERN, 376 (2000).
\bibitem{dark_usr} F. L. Pimpec et al., ``Dark Current Studies For SWISSFEL,'' arXiv:1205.3098v1 [physics.acc-ph] May 2012, \texttt{http://arxiv.org/abs/1205.3098v1}.

\end{thebibliography}
\end{document}